\PassOptionsToPackage{unicode}{hyperref}
\PassOptionsToPackage{hyphens}{url}
\PassOptionsToPackage{dvipsnames,svgnames,x11names}{xcolor}
\documentclass[
  12pt]{article}

\usepackage{amsmath,amssymb}
\usepackage{iftex}
\ifPDFTeX
  \usepackage[T1]{fontenc}
  \usepackage[utf8]{inputenc}
  \usepackage{textcomp} 
\else 
  \usepackage{unicode-math}
  \defaultfontfeatures{Scale=MatchLowercase}
  \defaultfontfeatures[\rmfamily]{Ligatures=TeX,Scale=1}
\fi
\usepackage{lmodern}
\ifPDFTeX\else  
\fi
\IfFileExists{upquote.sty}{\usepackage{upquote}}{}
\IfFileExists{microtype.sty}{
  \usepackage[]{microtype}
  \UseMicrotypeSet[protrusion]{basicmath} 
}{}
\makeatletter
\@ifundefined{KOMAClassName}{
  \IfFileExists{parskip.sty}{%
    \usepackage{parskip}
  }{
    \setlength{\parindent}{0pt}
    \setlength{\parskip}{6pt plus 2pt minus 1pt}}
}{
  \KOMAoptions{parskip=half}}
\makeatother
\usepackage{xcolor}
\makeatletter
\ifx\paragraph\undefined\else
  \let\oldparagraph\paragraph
  \renewcommand{\paragraph}{
    \@ifstar
      \xxxParagraphStar
      \xxxParagraphNoStar
  }
  \newcommand{\xxxParagraphStar}[1]{\oldparagraph*{#1}\mbox{}}
  \newcommand{\xxxParagraphNoStar}[1]{\oldparagraph{#1}\mbox{}}
\fi
\ifx\subparagraph\undefined\else
  \let\oldsubparagraph\subparagraph
  \renewcommand{\subparagraph}{
    \@ifstar
      \xxxSubParagraphStar
      \xxxSubParagraphNoStar
  }
  \newcommand{\xxxSubParagraphStar}[1]{\oldsubparagraph*{#1}\mbox{}}
  \newcommand{\xxxSubParagraphNoStar}[1]{\oldsubparagraph{#1}\mbox{}}
\fi
\makeatother

\usepackage{longtable,booktabs,array}
\usepackage{calc} 
\newcommand{\T}{\intercal}
\usepackage{etoolbox}
\makeatletter
\patchcmd\longtable{\par}{\if@noskipsec\mbox{}\fi\par}{}{}
\makeatother
\IfFileExists{footnotehyper.sty}{\usepackage{footnotehyper}}{\usepackage{footnote}}
\makesavenoteenv{longtable}
\usepackage{graphicx}
\makeatletter
\def\maxwidth{\ifdim\Gin@nat@width>\linewidth\linewidth\else\Gin@nat@width\fi}
\def\maxheight{\ifdim\Gin@nat@height>\textheight\textheight\else\Gin@nat@height\fi}
\makeatother
\setkeys{Gin}{width=\maxwidth,height=\maxheight,keepaspectratio}
\makeatletter
\def\fps@figure{htbp}
\makeatother

\makeatletter
\@ifpackageloaded{caption}{}{\usepackage{caption}}
\AtBeginDocument{%
\ifdefined\contentsname
  \renewcommand*\contentsname{Table of contents}
\else
  \newcommand\contentsname{Table of contents}
\fi
\ifdefined\listfigurename
  \renewcommand*\listfigurename{List of Figures}
\else
  \newcommand\listfigurename{List of Figures}
\fi
\ifdefined\listtablename
  \renewcommand*\listtablename{List of Tables}
\else
  \newcommand\listtablename{List of Tables}
\fi
\ifdefined\figurename
  \renewcommand*\figurename{Figure}
\else
  \newcommand\figurename{Figure}
\fi
\ifdefined\tablename
  \renewcommand*\tablename{Table}
\else
  \newcommand\tablename{Table}
\fi
}
\@ifpackageloaded{float}{}{\usepackage{float}}
\floatstyle{ruled}
\@ifundefined{c@chapter}{\newfloat{codelisting}{h}{lop}}{\newfloat{codelisting}{h}{lop}[chapter]}
\floatname{codelisting}{Listing}

\makeatother
\makeatletter
\@ifpackageloaded{caption}{}{\usepackage{caption}}
\@ifpackageloaded{subcaption}{}{\usepackage{subcaption}}
\makeatother
\ifLuaTeX
  \usepackage{selnolig}  
\fi
\usepackage[]{natbib}
\usepackage{bookmark}
\usepackage{dsfont}
\usepackage{bm}

\IfFileExists{xurl.sty}{\usepackage{xurl}}{} 
\usepackage[capitalize,noabbrev]{cleveref}
\hypersetup{
  pdftitle={mytitle},
  pdfauthor={Author 1; Author 2},
  pdfkeywords={3 to 6 keywords, that do not appear in the title},
  colorlinks=true,
  linkcolor={blue},
  filecolor={Maroon},
  citecolor={Blue},
  urlcolor={Blue},
  pdfcreator={LaTeX via pandoc}}

\newcommand{\anon}{1}

\begin{document}

\def\spacingset#1{\renewcommand{\baselinestretch}%
{#1}\small\normalsize} \spacingset{1}


\if1\anon
{
  \title{\bf A nutritionally informed model for Bayesian variable selection with metabolite response variables}
  \author{Dylan Clark-Boucher\\
    Department of Biostatistics, Harvard TH Chan School of Public Health\\
    Brent A Coull\thanks{B.A.C. is supported by the National Institute of Environmental Health Sciences (P30ES000002)}\\
    Department of Biostatistics, Harvard TH Chan School of Public Health\\
    Harrison T Reeder \\
    Biostatistics, Massachusetts General Hospital \\ 
    Fenglei Wang \\
    Department of Nutrition, Harvard TH Chan School of Public Health\\
    Qi Sun \\
    Department of Nutrition, Harvard TH Chan School of Public Health\\   
    Jacqueline R Starr\textsuperscript{\textdagger\ddag}  \\
    Channing Division of Network Medicine, Brigham and Women's Hospital\\
    Kyu Ha Lee \thanks{J.R.S. and K.L. are supported by the National Institute of Dental and Craniofacial Research (R03DE027486) and the National Institute of General Medical Sciences (R01GM126257)} \footnote{Co-senior authors} \\
    Departments of Nutrition, Harvard TH Chan School of Public Health    }
  \maketitle
} \fi

\if0\anon
{
  \bigskip
  \bigskip
  \bigskip
  \begin{center}
    {\LARGE\bf A nutritionally informed model for Bayesian variable selection with metabolite response variables}
\end{center}
  \medskip
} \fi

\bigskip
\begin{abstract}
Understanding the pathways through which diet affects human metabolism is a central task in nutritional epidemiology. This article proposes novel methodology to identify food items associated with blood metabolites in two cohorts of healthcare professionals. We analyze 30 food intake variables that exhibit relationship structure through their correlations and nutritional attributes. The metabolic responses include 244 compounds measured by mass spectrometry, presenting substantial challenges that include missingness, left-censoring, and skewness. While existing methods can address such factors in low-dimensional settings, they are not designed for high-dimensional regression involving strongly correlated predictors and non-normal outcomes. To address these challenges, we propose a novel Bayesian variable selection framework for metabolite response variables based on a skew-normal censored mixture model. To exploit substantive information on the nutritional similarities among dietary factors, we employ a Markov random field prior that encourages joint selection of related predictors, while introducing a new, efficient strategy for its hyperparameter specification. Applying this methodology to the cohort data identifies multiple metabolite-diet associations that are consistent with previous research as well as several potentially novel associations that were not detected using standard methods. The proposed approach is implemented in the  R package \texttt{multimetab}, facilitating its use in high-dimensional metabolomic analyses.
\end{abstract}

\noindent%
\vfill

\newpage
\spacingset{1.8} 

\section{Introduction}\label{sec-intro}

A central goal of metabolomics is to understand how age, sex, health, nutrition, and other factors affect the abundance of blood metabolites. Metabolites are small molecules produced or expended by metabolism---an umbrella term for the chemical reactions that keep the human body alive and functioning. Metabolites play essential roles in processes such as signaling, immune modulation, and energy production, among others \citep{Baker_Rutter_2023}. Most metabolites are derived from digested compounds or produced as a chemical byproduct of the gut microbiome, making nutritional intake an important driver of their abundance levels \citep{Jia_Yang_Wilson_Mueller_Sears_Treisman_Robinson_2022, Caspani_Swann_2019}. Quantifying the relationships between nutritional factors and metabolite abundance is essential for elucidating the metabolic pathways that underpin human health. 

The objective of this article is to identify dietary predictors of metabolites using data from two cohort studies of healthcare professionals: the Mind Body Study (MBS) and the Men's Lifestyle Validation Study (MLVS) \citep{Huang_Trudel-Fitzgerald_Poole_Sawyer_Kubzansky_Hankinson_Okereke_Tworoger_2019, Wang_Nguyen_Li_Yan_Ma_Rinott_Ivey_Shai_Willett_Hu_etal._2021}. Recent analyses of MBS and MLVS have established relationships between specific types of metabolites and diet, cardiometabolic risk, and the gut microbiome \citep{Li_Li_Ivey_Wang_Wilkinson_Franke_Lee_Chan_Huttenhower_Hu_etal._2022, Hamaya_Sun_Li_Yun_Wang_Curhan_Huang_Manson_Willett_Rimm_etal._2024, Malik_Guasch-Ferre_Hu_Townsend_Zeleznik_Eliassen_Tworoger_Karlson_Costenbader_Ascherio_etal._2019, Li_Wang_Li_Ivey_Wilkinson_Wang_Li_Liu_Eliassen_Chan_etal._2022}. These analyses have mainly been narrow in scope, targeting a subset of the measured metabolites and considering only a few food intake variables, or aggregated versions of them, individually. We expand on this research by performing a comprehensive scan over 244 metabolites of diverse types. Each metabolite is treated as the response variable in a separate regression model that adjusts for all 30 food intake variables jointly, allowing us to pinpoint specific food items that are most informative. We accomplish this by developing a novel Bayesian variable selection framework for metabolite responses that exploits external information on the nutritional similarity of the considered food items. 

\subsection{Application challenges}
\label{sec:challenges}
\begin{figure}[ht!]
    \centering
    \includegraphics[width=6.5in]{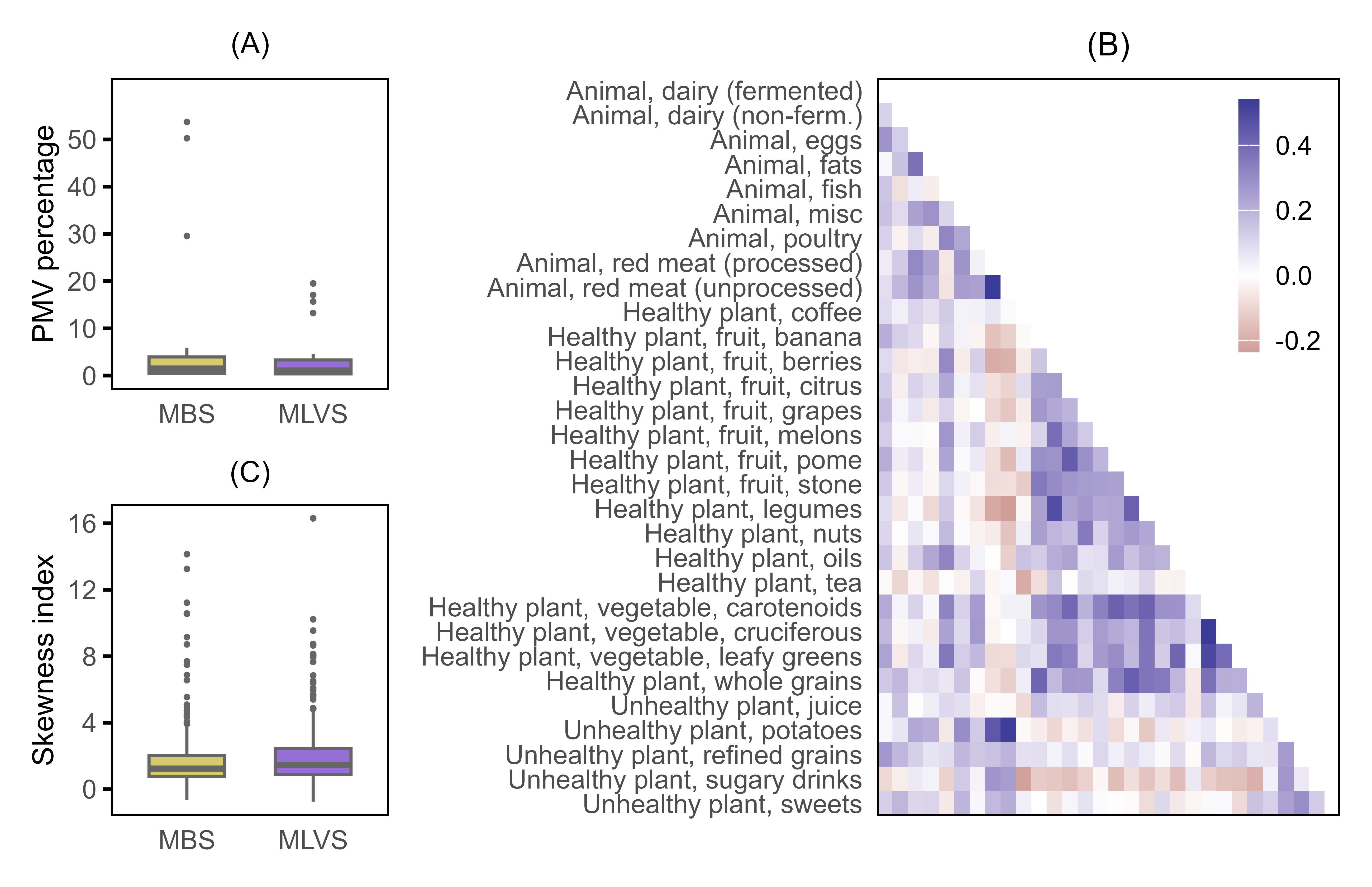} 
    \caption{\textbf{MBS and MLVS data challenges}. (A) The point mass value (PMV) percentages of 30 metabolites with at least one PMV in the Mind Body Study (MBS) and Men's Lifestyle Validation Study (MLVS). (B) The skewness index of all 244 metabolites, where more positive values indicate stronger right skewness. (C) Pairwise correlations among the 30 dietary intake predictor variables after combining the two datasets. We used half-minimum imputation for panels (B) and (C).}
    \label{fig:data}
\end{figure}

In common with most metabolomics datasets, the considered data present an array of challenges for statistical analysis. First, mass spectrometry-based metabolomics data exhibit a specific type of missing values termed point mass values (PMVs) \citep{Hrydziuszko_Viant_2012, Do_Wahl_Raffler_Molnos_Laimighofer_Adamski_Suhre_Strauch_Peters_Gieger_etal._2018}. ``Biological'' PMVs arise when the mass spectrometer fails to detect a metabolite because it is completely absent from the sample. ``Technical'' PMVs occur when a metabolite's intensity is below a threshold called the ``detection limit,'' producing a spurious missing value that reflects censoring, not true absence. It is generally unknown whether individual PMVs are technical or biological \citep{Taylor_Leiserowitz_Kim_2013}. In the MLVS and MBS data, 12\% of metabolites have at least one PMV. The metabolite missing a value the most often has 54\% PMVs (\cref{fig:data}A).

Second, the dietary predictors exhibit complex dependence structure through their correlations and nutritional relationships. The Pearson correlations among the 30 food intake variables ranged from -0.24 to 0.54 (\cref{fig:data}B), requiring advanced statistical techniques, such as variable selection, that can extract sparse signals from complex feature spaces. Adding to the complexity, some pairs of foods have similar metabolic signatures due to their source or nutritional content, such as types of fruit. \citet{Satija_Bhupathiraju_Rimm_Spiegelman_Chiuve_Borgi_Willett_Manson_Sun_Hu_2016} exploited such information to identify dietary influences of type II diabetes using cohort data overlapping with MBS and MLVS, grouping foods by their nutritional value to calculate healthy and unhealthy diet scores. An ideal modeling framework should similarly use nutritional relationships to promote powerful, biologically informed inference.

Finally, metabolites tend to be strongly right skewed and have many outliers. Excessive skewness can invalidate statistical methods that assume Gaussian errors, introducing bias and distorting inference \citep{Wilson_Loughran_Brame_2020}. The metabolites in MBS and MLVS have a mean skewness index of 2.3, indicating heavy right skewness (\cref{fig:data}C). 

\subsection{Existing methods and proposed framework}

There are two general frameworks for handling PMVs in statistical analysis. One strategy is to use imputation methods designed specifically for mass spectrometry data, which are essential because PMVs are missing-not-at-random, invalidating standard imputation tools \citep{Wei_Wang_Su_Jia_Chen_Chen_Ni_2018}. The second strategy, adopted in this article, is to use advanced statistical methods that handle PMVs internally \citep{Taylor_Leiserowitz_Kim_2013, Gleiss_Dakna_Mischak_Heinze_2015, Huang_Wang_2022}.  The simplest approach of this type is to perform one-part tests based on tobit or accelerated failure time (AFT) models, assuming the PMVs are technical PMVs caused by censoring. In contrast, two-part models compare the PMVs and non-PMVs separately and combine the p-values, treating the PMVs as biological. To flexibly account for PMVs of both types, hybrid approaches---such as tobit or AFT mixture models---have been developed. Recent extensions of this framework have used Bayesian modeling and empirical Bayes shrinkage techniques \citep{Shah_Brock_Gaskins_2019,Huang_Lane_Fan_Higashi_Weiss_Yin_Wang_2020}. 

Although mixture models are well established for accommodating PMVs in metabolomics studies, their existing variants are unsuitable for handling correlated and nutritionally related dietary factors as predictors. Specifically, no existing mixture models offer frameworks for performing variable selection, rendering them inadequate for detecting signals obscured by correlated, high-dimensional sets of covariates. Nor do they offer architecture for using external information on the substantive relationships among predictors, meaning the nutritional patterns of the food intake variables cannot be exploited. 

Existing mixture models also rely on ad hoc strategies for handling skewness, such as assuming the log-transformed metabolites have Gaussian errors.  The main weakness of transformations is that some metabolites are still highly skewed after being transformed. Though distributional violations are usually inconsequential for standard linear regression models, they can greatly impact the performance of censored data methods such as tobit models \citep{Wilson_Loughran_Brame_2020}. A preferable strategy might be to incorporate skewness estimation into the model framework, adapting to the specific skewness of each metabolite. Lastly, though not essential, it is helpful to have the option to analyze untransformed data to preserve the original interpretation of the regression coefficients.

We address these challenges by developing an innovative Bayesian variable selection framework that: (1) internally accounts for technical and biological PMVs; (2) identifies predictive covariates within complex feature spaces; (3) exploits structural information on the substantive relationships among covariates; and (4) directly models the skewness of each metabolite. The proposed framework assumes a skew-normal censored mixture (SNCM) model, flexibly handling missingness and skewness without requiring imputation or transformation.  To identify dietary factors associated with each metabolite, we perform Bayesian variable selection via spike-and-slab priors. To exploit the nutritional relationships among dietary factors, we adopt a Markov random field (MRF) hyperprior that promotes the joint selection of nutritionally similar variables \citep{Stingo_Chen_Tadesse_Vannucci_2011}. Recognizing the computational burden of standard MRF implementations, which often require repeated model fitting for hyperparameter tuning, we propose an efficient, data-independent strategy for hyperparameter specification. We provide user-friendly software in the R package \texttt{multimetab}.

\section{Model and Posterior Inference}
\subsection{Model formulation}
For $i\in1,\dots,n$, let $Y_i$ denote the observed value of the $i\textsuperscript{th}$ subject's metabolite. Let $Y_i^{*}$ be the corresponding ``true'' value that is observed only if $Y_i^{*}\geq\psi$, where $\psi>0$ is the detection limit that induces censoring. Let $\bm{X}_i=(X_{1i},\dots,X_{pi})^{\T}$ denote a possibly high-dimensional vector of $p$ covariates, some associated with $Y_i^{*}$ and others not. The variables in $\bm{X}_i$ are the main covariates of interest and will undergo variable selection.  Let  $\bm{C}_i=(C_{1i},\dots,C_{si})^{\T}$ denote a low-dimensional vector of $s$ covariates, $s<<n$, that are excluded from variable selection because they are identified \emph{a priori} as potential confounding variables. For consistency of language, we will refer to $X_{1i},\dots,X_{pi}$ as ``predictors'' and $C_{1i},\dots,C_{si}$ as ``confounders.'' Let $U_i{\sim} \text{Bernoulli}(\rho)$ be indicator variables marking whether the metabolite is present in the sample or absent, and let $V_i$ be the value of the metabolite when it is present. We adopt a hierarchical model:
\begin{gather*}
    V_i = \beta_0+\sum_{j=1}^p\beta_jX_{ij}+ \sum_{t=1}^s\alpha_tC_{it} + \epsilon_i\\
    Y_i^{*} = U_iV_i\\
    Y_i = \begin{cases} 
      Y^{*}_i & Y^{*}_i \geq\psi \\
      \text{PMV} & Y^{*}_i <\psi \\
   \end{cases}
\end{gather*}
The PMVs, where $Y_i$ is missing, have either $U_i=0$, $V_i<\psi$, or both. PMVs with $U_i=0$ denote biological PMVs caused by the total absence of the metabolite from the sample, whereas PMVs with $U_i=1$ and $V_i<\psi$ denote technical PMVs caused by censoring. We treat $V_i$ as the response variable while assuming $\epsilon_i\overset{iid}{\sim}\text{SN}(0, \sigma^2,\delta)$, a skew-normal distribution with variance parameter $\sigma^2$ and skewness $\delta$ under the parameterization given by \citet{Sahu_Dey_Branco_2003}. This is equivalent to assuming the error distribution is $\text{N}(0,\sigma^2)+\delta|\text{N}(0,1)|$, with mean $\delta\sqrt{2/\pi}$ and variance $\sigma^2+(1-\sqrt{2/\pi})\delta^2$. Our main interest lies in $\bm{\beta}=(\beta_{1},\dots,\beta_{p})^{\T}$, which contains the linear associations between the predictors and the metabolite among samples where the metabolite is present (i.e., those with $U_i=0)$. The vector $\bm{\alpha}=(\alpha_{1},\dots,\alpha_{s})^{\T}$ denotes the associations corresponding to the confounding variables.  

Let $F_\text{SN}$ denote the cumulative distribution of the skew-normal distribution and $f_\text{SN}$ the corresponding density function. Let $W_i=\mathds{1}(Y_i^{*}\geq \psi)$ be an indicator variable for whether $Y_i$ is observed (i.e., not a PMV), and define $\mu_i=\beta_0 + \bm{X}_i^{\T}\bm{\beta} + \bm{C}_i^{\T} \bm{\alpha} $. Then, letting $\bm{\theta}=(\beta_0,\bm{\beta},\bm{\alpha},\sigma^2,\delta,\rho)$, the likelihood for observation $i$ is
\begin{gather*}
    \mathcal{L}_i(\bm{\theta};y_i,\bm{x}_i,\bm{c}_i) = \biggl(\rho f_\text{SN}(y_i;\mu_i ,\sigma,\delta)\biggr)^{w_i}\biggl(1-\rho + \rho F_\text{SN}(\psi;\mu_i ,\sigma,\delta)\biggr)^{1-w_i}.
\end{gather*}

\subsection{Prior specification and incorporation of external information}
To identify predictors associated with the abundance of a given metabolite, we place variable selection priors on the entries in $\bm{\beta}$. Specifically, letting $\gamma_j$ be a latent indicator variable for whether $\beta_j\ne0$, $j\in1,\dots,p$, we apply the spike-and-slab priors $\beta_j^{*}\sim_{iid}\text{N}(0,\nu^2)$, $\beta_j=\beta^{*}\gamma_j$. While it is more common to write such priors in the simpler form $\beta_j\sim\text{N}(0,\gamma_j\nu^2)$ \citep{Brown_Vannucci_Fearn_1998}, omitting the latent variable $\beta_j^{*}$, we use the former version because it allows the parameters be updated by using Gibbs sampling \citep{Kuo_Mallick_1998}.

In Bayesian variable selection methods based on spike-and-slab priors, $\gamma_1,\dots,\gamma_p$ are typically assigned independent Bernoulli hyperiors that allot equal inclusion probability to all predictors. However, many of the dietary predictors in the motivating data share nutritional components that may inform their metabolic effects, for example, fermented and non-fermented dairy products. Yet, collapsing such categories risks obscuring any differences between them and may reduce sensitivity to detect their metabolic signatures. To exploit such structure in the variable selection scheme, we adopt a structured multivariate prior, the MRF prior \citep{Li_Zhang_2010, Zhao_Banterle_Lewin_Zucknick_2024}, on the vector $\bm{\gamma}=(\gamma_1,\dots,\gamma_p)^{\T}$. 

Let $\bm{R}=[r_{j,j^\prime}]\in\mathbb{R}^{p\times p}$ denote a symmetric matrix with zero diagonal and non-negative off-diagonal entries, where $r_{j,j^\prime}$ quantifies the strength of the relationship between predictors $X_{j}$ and $X_{j^\prime}$. The form of the MRF prior is 
\begin{gather*}
    P(\bm{\gamma})\propto \exp\biggl(\omega \sum_{j=1}^p\gamma_j + \eta\bm{\gamma}^{\T} \bm{R}\bm{\gamma}\biggr),
\end{gather*}
where $\omega\in\mathbb{R}$ and $\eta\geq0$ are hyperparameters controlling the model's sparsity such that higher values tend to produce models with more variables. $\omega$ controls overall sparsity uniformly across variables, and $\eta$ regulates the influence of the relationship structure given by $\bm{R}$. When $\eta=0$, the prior reduces to an independent Bernoulli prior with inclusion probability equal to $\text{logit}^{-1}(\omega)$. When $\eta>0$, the prior encourages the joint selection of variables that are mutually related, such that greater values of $r_{j,j^\prime}$ increase the probability of selecting $X_{j}$ once $X_{j^\prime}$ is selected. Greater values of $\eta$ enhance this effect.

For the remaining parameters, we assume $\beta_0\sim \text{N}(0,\nu_\text{0}^2)$, $\sigma^2\sim \text{Inverse-Gamma}(0.5\xi,0.5\xi_0\sigma_0^2)$, $\delta\sim \text{N}(0,\nu_\text{d}^2)$, $\alpha_t\sim \text{N}(0,\lambda_t^2)$, $t\in1,\dots,s$, and $\rho\sim \text{Beta}(\rho_0,\rho_1)$. These priors reduce to conjugate priors after the introduction of additional latent variables, discussed next. 

\subsection{Posterior inference and computation}
\label{sec:inference}
 We perform inference via Markov chain Monte Carlo (MCMC) using Gibbs sampling. The specified prior distributions can be viewed as conjugate priors after introducing latent variables to handle non-standard components of the model. First, we decompose the skew-normal error term into a normal and a half-normal component as suggested by \citet{Sahu_Dey_Branco_2003}. Specifically, an equivalent definition of $V_i$ is 
\begin{gather}
\label{eq:latent_model}
        V_i = \beta_0 + \sum_{j=1}^p\gamma_j\beta_j^{*}X_{ji} + \sum_{t=1}^s\alpha_tC_{ti}+\delta Z_i + \zeta_i,
\end{gather}
$i\in1,\dots,n$, where $Z_i\overset{iid}{\sim}|\text{N}(0,1)|$ and $\zeta_i\overset{iid}{\sim} \text{N}(0,\sigma^2)$. Conditional on the latent variables $\bm{\gamma}$, $\bm{V} =(V_{1},\dots, V_n)^{\T}$, and $\bm{Z} = (Z_1,\dots, Z_n)^{\T}$, the model reduces to a standard linear regression model with Gaussian errors, for which the priors on $\beta_0$, $\sigma^2$, $\delta$, $\bm{\beta}^{*}$, and $\bm{\alpha}$ are conjugate. Likewise, the prior on $\rho$ is conjugate conditional on the latent variables $\bm{U} = (U_1,\dots,U_n)^{\T}$. 

We use data augmentation to introduce $\bm{Z}$, $\bm{V}$, and $\bm{U}$ to the inferential scheme \citep{Albert_Chib_1993}. Each $Z_i$ is updated based on its full conditional distribution assuming a half-normal conjugate prior. Each $V_i$ equals $Y_i$ if $W_i=1$ (non-PMVs), follows a $\text{N}(\mu_i+\delta Z_i,\sigma^2)$ distribution if $W_i=0$ and $U_i=0$, and follows a $\text{N}(\mu_i+\delta Z_i,\sigma^2)\mathds{1}(-\infty, \psi)$ distribution if $W_i=0$ and $U_i=1$. Finally, each $U_i$ equals 1 if $W_i=1$, equals 0 if $W_i=0$ and $V_i > \psi$, and follows a $\text{Bernoulli}(\rho)$ distribution if $W_i=0$ and $V_i <\psi$. The augmented likelihood in terms of latent variables is
\begin{gather*}
\mathcal{L}_i(\bm{\theta};V_i,U_i,Z_i,\bm{X}_i,\bm{C}_i)=(1-\rho)^{1-U_i}\rho^{U_i}\frac{2}{\delta\sigma}\phi\biggl(\frac{Z_i}{\delta}\biggr)\phi\biggl(\frac{v_i-\mu_i-\delta Z_i}{\sigma}\biggr)\mathds{1}(Z_i>0),
\end{gather*}
where $\phi(\cdot)$ is the standard normal density. 

The convenient structure of the \citet{Kuo_Mallick_1998} spike-and-slab prior allows $\bm{\beta}^{*}$ and $\bm{\alpha}$ to be updated by their full conditional distribution. The $\beta_j^{*}$ parameters for which $\gamma_j=1$ are updated jointly with $\bm{\alpha}$ using the standard update for Bayesian linear regression. If $\gamma_j=0$, then $\beta_j^{*}$ does not affect the likelihood and can be updated by sampling directly from the prior. Finally, we update each $\gamma_j$ by its full conditional, exploiting the fact that $\gamma_j\mid \{\gamma_{j^\prime}\}_{j\prime\ne j}$ has a tractable Bernoulli prior where
\begin{align*}
    \text{Pr}(\gamma_{j}=1\mid \{\gamma_{j\prime}\}_{j\prime\ne j})= \text{logit}^{-1}\biggl(\omega + 2\eta\sum_{j^\prime\ne j} r_{j,j\prime}\gamma_{j^\prime}\biggr).
\end{align*}
Further details are provided in the supplement (Section 1).

\subsubsection{A prior-based method for MRF hyperparameter specification}
\label{sec:determine_eta}
An important consideration when implementing the MRF prior is specifying $\eta$, which controls the degree to which the relationship matrix, $\bm{R}$, affects variable selection. Prior work has shown that MRF-based priors can exhibit `phase transition' behavior: as $\eta$ increases beyond a critical threshold, the prior can shift abruptly from favoring sparse models to selecting nearly all variables \citep{Li_Zhang_2010, Zhao_Banterle_Lewin_Zucknick_2024}. This behavior may lead to unstable posterior inference and obscure the interpretability of identified associations.

Several approaches have been proposed to address this challenge. One option is to place a hyperior on $\eta$ to infer it directly \citep{Stingo_Chen_Tadesse_Vannucci_2011}. Another is to conduct a data-driven grid search to find the largest $\eta$ value below the phase transition point \citep{Li_Zhang_2010, Lee_Tadesse_Baccarelli_Schwartz_Coull_2017}. However, both strategies require exploring large $\eta$ values that can produce highly parameterized models, often creating a substantial computational burden. 

To circumvent these issues, we introduce a new approach for choosing $\eta$ that depends only on $\omega$ and $\bm{R}$. The idea is to curtail the prior probability of a large model by finding an MRF prior with summary statistics similar to those from a reasonable independent prior. The first step is to select $\omega$  based on a target baseline sparsity level. For example, choosing $\omega=\text{logit}(0.01)$ corresponds to a 1\% prior inclusion rate under independent Bernoulli hyperpriors ($\eta=0$). After setting $\omega=\omega_0$, we choose the largest $\eta$ such that the prior $95\textsuperscript{th}$ percentile model size is below that of the MRF$(\omega=\text{logit}(2\text{logit}^{-1}(\omega_0)), \eta=0)$, an independent Bernoulli prior with inclusion probability $2\omega_0$. We search for such an $\eta$ over a list of candidates by sampling from the corresponding prior distributions, which is substantially more computationally efficient than existing posterior- or data-driven tuning methods.

\section{Simulation Studies}
We evaluated the performance of the proposed inferential scheme through a series of simulation studies. The goal was to assess its robustness under varying levels of skewness, censoring, residual variance, and high-dimensionality---challenges commonly encountered in mass spectrometry-based metabolomics data.

\subsection{Data generation and settings}
\label{sec:data_gen}
For a baseline setting, we generated data under the SNCM model with $n=400$ observations, $p=300$ predictors, and no confounding variables ($s=0$). We set $\beta_0=5$ and generated predictors as $\bm{X}_i\overset{iid}{\sim}\text{MVN}(0,\bm{I})$. We chose $\sigma$ and $\delta$ so that the residual variance in $V_i$ was fixed at 8, with 75\% explained by the half-normal latent variable $Z_i$. To induce an average censoring rate of 36\%, we set $\rho=0.8$ and chose $\psi$ such that $P(V_i<\psi)=0.20$. 

\begin{figure}[htbp]
    \centering
    \includegraphics[width=5in]{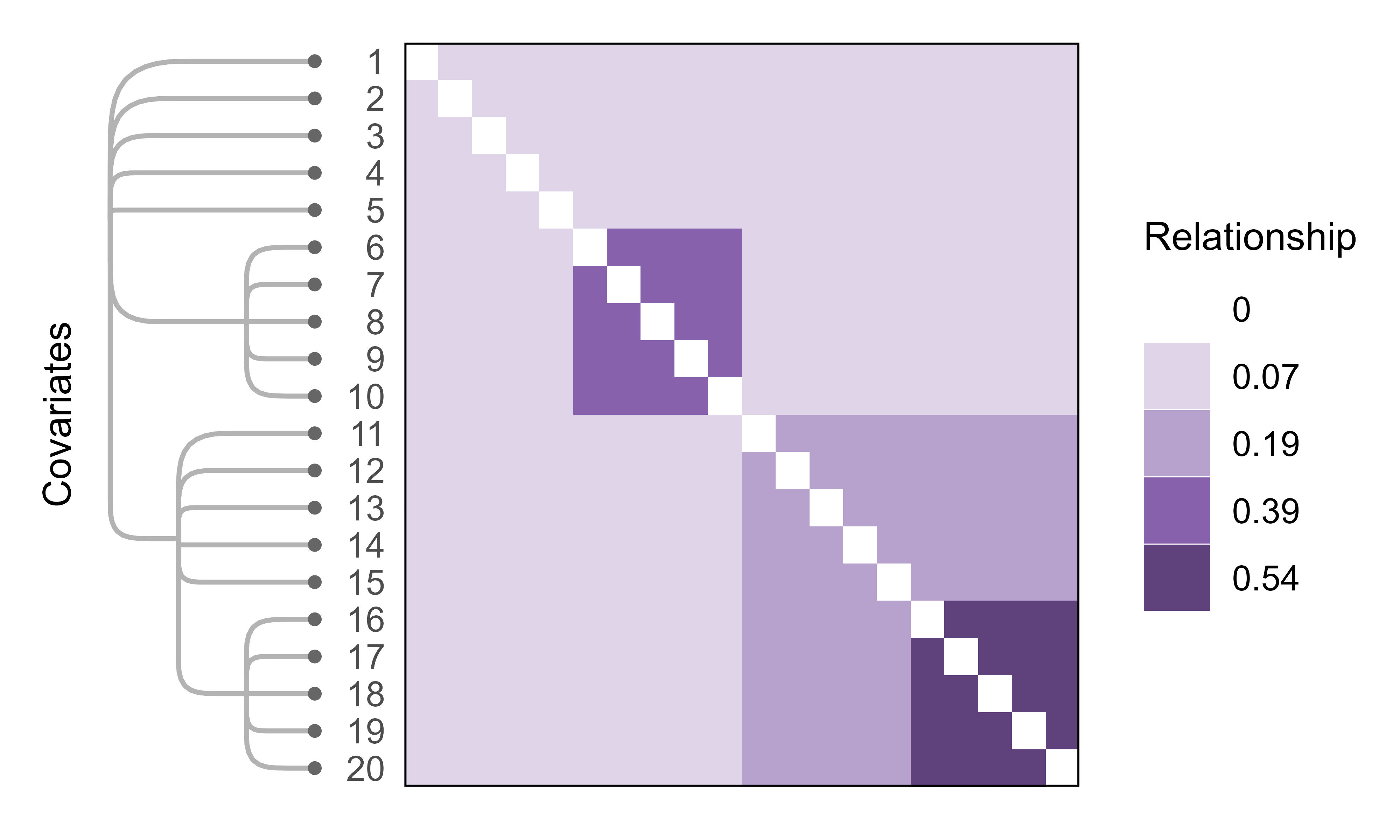} 
    \caption{\textbf{Relationships within a block of 20 predictors in the simulation design}. The heatmap depicts one of 15 identical predictor blocks in the relationship matrix $\bm{R}$. Predictors within the same block have structured relationships based on the hierarchical tree shown on the left, and predictors in different blocks are unrelated.}
    \label{fig:simulation_R}
\end{figure}

To mimic the hierarchical dietary relationships in the MLVS/MBS data (\cref{sec:diet_R}), we defined $\bm{R}\in\mathbb{R}^{300\times300}$ as a block-diagonal matrix with 15 blocks (categories) of 20 connected predictors.  Within each category, predictors 6-10 and 11-20 formed two subcategories, and predictors 16-20 formed a sub-subcategory  (\cref{fig:simulation_R}). For predictors $j$ and $j^\prime$, we let $r_{j,j^\prime}=\exp(a_{j,j^\prime}/c)/b$, where $a_{j,j^\prime}\in\{1,2,3\}$ is the depth of their deepest shared category/subcategory; $b_{j,j^\prime}\in\{1,\dots,20\}$ is the size of their deepest shared category/subcategory; and $c=3$ is the maximum depth in the hierarchy. For example, since predictors 15 and 16 are in a subcategory ($a=2$) of size 10, their relationship strength is $\exp(2/3)/10\approx 0.19$. Since predictors 16 and 17 are in a sub-subcategory ($a=3$) of size 5, their relationship is $\exp(3/3)/5\approx 0.54$. Predictors in different blocks have a relationship strength of 0. 

The construction of $\bm{\beta}\in\mathbb{R}^{300}$ was motivated by two goals: to induce variety in the signal strengths and directions among predictors, and to create correspondence with the relationship structure defined in $\bm{R}$. Specifically,  in each of the first four predictor blocks/categories used to construct $\bm{R}$, we assigned five predictors to be associated with the response (20 total). We chose predictors 1-5 (weakly connected through $\bm{R}$) in block 1, 6-10 (highly connected) in block 2, 11-15 (moderately connected) in block 3, and 16-20 (very highly connected) in block 4. The regression coefficients within each block were set to $\{0.4, -0.6, 0.8, -1.0,1.2\}$, varying the signal strength and direction. 

To evaluate performance under varied conditions, we considered seven additional data-generating scenarios in each of which one aspect of the baseline setting was perturbed. (i) \emph{Higher noise} was induced by multiplying both $\sigma$ and $\delta$ by 1.5, which raised the residual variance of $V_i$ to 18; (ii) \emph{Heavier censoring} was created by setting $\rho=0.7$ and $P(V_i<\psi)=0.30$, resulting in an overall censoring rate of 51\%; (iii) \emph{Greater skewness} was obtained by adjusting $\sigma$ and $\delta$ so that the half-normal latent variable explained 95\% (instead of 75\%) of the error variance while keeping total variance unchanged; (iv) \emph{Misspecified dependence structure} was simulated by randomly  permuting the rows and columns of $\bm{R}$ in each replicate, thereby breaking the link between relationships among predictors and their associations with the response; (v) \emph{Non-normal errors} were examined by drawing $\epsilon_i$ from a log-normal distribution, matched in mean and variance to the skew-normal error used in the baseline scenario; (vi) \emph{Correlated predictors} were introduced by sampling $\bm{X}_i\overset{iid}{\sim}\text{MVN}(0,\bm{\Sigma})$, where $\bm{\Sigma}$ was obtained by replacing the diagonal of $\bm{R}$ with ones; and (vii) \emph{Large sample size} was considered by raising $n$ to 1,000. The setting with correlated predictors was intended to be directly relevant to the proposed analysis of correlated food intake variables.

\subsection{Analysis and hyperparameters}
\label{sec:sim_analysis}
We fit the SNCM model to 800 simulated replicates of each setting and summarized performance in terms of the mean overall true positive rate (TPR), variable-specific TPR, and false discovery rate (FDR). Letting $\tilde\gamma_j$ be an indicator variable for whether predictor $j$ is selected for a given dataset, the variable-specific true positive rate is $\mathbb{E}[
\tilde\gamma_j\mid \gamma_j=1]$, the mean overall TPR is $\mathbb{E}[\sum_j\gamma_j\tilde\gamma_j / \sum_j\gamma_j\mid\bm{\gamma}]$, and the FDR is $\mathbb{E}[\sum_j(1-\gamma_j)\tilde\gamma_j / \sum_j\tilde\gamma_j\mid\bm{\gamma}]$.
We also calculated the variable-specific bias and root mean squared error (rMSE) for estimating the regression coefficients of the effective predictors. Hyperparameters were fixed at $\nu^2_0=5^2$, $\nu^2_d=5^2$, $\nu^2=2^2$, $\xi_0=5$, $\sigma^2_0=4$, and $\omega=\text{logit}(0.02)$. We searched for $\eta$ over the range $0.01r^{-1},0.02r^{-1},\dots,1r^{-1}$, where $r=\text{max}(\bm{R})$. We adapted $\rho_0$ and $\rho_1$ to each data set by setting $\rho_0=5\sqrt{\bar{W}}$ and $\rho_1=5(1-\sqrt{\bar{W}})$, where $\bar{W}=\sum_{i=1}^nW_i/n$; this yields a weakly informative prior centered on $P(U_i=0)=P(V_i<\psi)$. Since the $\psi$ is unknown in practice, we estimated it from the data as $\hat\psi=\min\{Y_i:W_i=1\}$. 

For each simulation scenario, we assessed the performance of the SNCM model with either independent Bernoulli priors (setting $\eta=0$) or the MRF prior. We did not focus on comparing strategies for handling PMVs since the use of mixture models is well-established \citep{Taylor_Leiserowitz_Kim_2013, Huang_Wang_2022}. However, for completeness, we additionally benchmarked the SNCM model against two ad hoc treatments of PMVs for the baseline scenarios: (i) treating all PMVs as technical (i.e., forcing $\rho=1$), and (ii) half-minimum imputation \citep{Wei_Wang_Su_Jia_Chen_Chen_Ni_2018}. For both approaches we used independent Bernoulli priors. 

To analyze each dataset, we drew 125,000 posterior samples after a burn-in period of 25,000,  retaining only 1 in 25 samples to reduce autocorrelation. We considered variables to be selected by the model if their posterior inclusion probability (PIP), $\hat{\gamma}_j=\hat{\mathbb{E}}[\gamma_j\mid \bm{X}=\bm{x},\bm{Y}=\bm{y}]$, exceeded a threshold chosen to achieve a 5\% Bayesian FDR \citep{Newton_Noueiry_Deepayan_Paul_2004}, where $\bm{y}$ and $\bm{x}$ denote the observed realizations of $\bm{Y}$ and $\bm{X}$, respectively . We estimated the regression coefficients  as $\hat\beta_j =\hat{\mathbb{E}}[\beta_j\mid \bm{X}=\bm{x},\bm{Y},\gamma_j=1]$. 

\subsection{Results}
\begin{figure}[ht]
    \centering
    \includegraphics[width=6.5in]{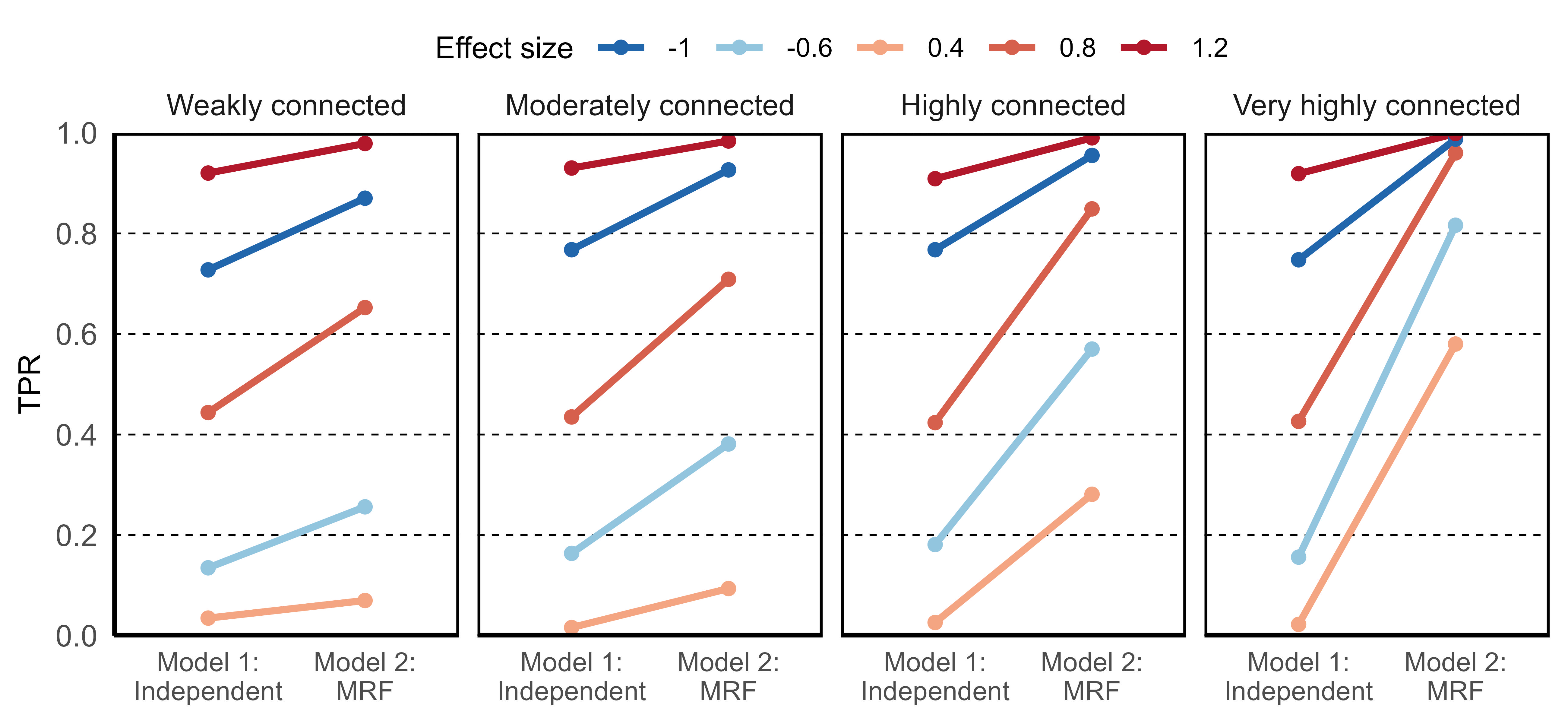} 
    \caption{\textbf{Variable-specific TPR among clusters of substantively connected variables in baseline simulations}. Each panel depicts true positive rate (TPR) of five predictors with different effect sizes in a given cluster.  Predictors in the same cluster are either weakly (0.07), moderately (0.19), highly (0.39), or very highly (0.54) connected through $\bm{R}$. Lines indicate the TPR change when using the Markov random field (MRF) prior to exploit relationship information instead of using independent Bernoulli priors.}
    \label{fig:variable_tpr}
\end{figure}

Under baseline simulation conditions, the SNCM model achieved uniformly higher TPR with the MRF prior than with the independent Bernoulli priors within all four relationship-strength clusters (\cref{fig:variable_tpr}). The improvement was relatively small for weakly connected predictor clusters and became increasingly pronounced in the moderately, highly, and very highly connected clusters. For example, within the very highly connected cluster the TPR for a predictor whose effect size was 0.4 increased from $<5$\% to nearly 60\%. Variable-specific bias remained negligible and similar across variables and priors, and the rMSE was slightly lower for the MRF prior (Supplementary Table 1). 

The above-described TPR gains were obtained without inflating the the FDR (\cref{tab:performance}). Across all the settings, FDR remained within 0.01--0.03 for the MRF prior and 0.01--0.04 for the independent Bernoulli priors, close to the 5\% nominal level. Even in more challenging settings---under higher variance, heavier censoring, increased skewness, correlated predictors, misspecified dependence, or misspecified errors---the MRF prior consistently outperformed the independent Bernoulli prior in TPRs with comparable FDRs. In contrast, treating all PMVs as technical or imputing them with the half-minimum rule led to substantially poorer TPR and FDR (Supplementary Table 2).

\begin{table}[ht]
\centering
\begin{tabular}{lcccc}
\toprule
\textbf{Data generation} & \multicolumn{2}{c}{\textbf{TPR}} & \multicolumn{2}{c}{\textbf{FDR}} \\
\cmidrule(lr){2-3} \cmidrule(lr){4-5}
 & \multicolumn{1}{c}{Independent} & \multicolumn{1}{c}{MRF} 
 & \multicolumn{1}{c}{Independent} & \multicolumn{1}{c}{MRF} \\
\midrule
Baseline & 0.53 (0.11) & 0.75 (0.09) & 0.01 (0.03) & 0.01 (0.03) \\ 
High Variance & 0.22 (0.10) & 0.39 (0.15) & 0.04 (0.11) & 0.03 (0.09) \\ 
High Censoring & 0.36 (0.12) & 0.60 (0.14) & 0.02 (0.06) & 0.02 (0.05) \\ 
High Skewness & 0.54 (0.11) & 0.77 (0.09) & 0.01 (0.03) & 0.01 (0.03) \\ 
Misspecified $\bm{R}$ & 0.53 (0.11) & 0.56 (0.11) & 0.01 (0.03) & 0.02 (0.04) \\ 
Log-normal Errors & 0.55 (0.11) & 0.77 (0.09) & 0.01 (0.03) & 0.01 (0.03) \\ 
Correlated Predictors & 0.38 (0.11) & 0.60 (0.14) & 0.02 (0.05) & 0.02 (0.05) \\ 
Large $n$ & 0.87 (0.06) & 0.95 (0.04) & 0.00 (0.01) & 0.01 (0.02) \\ 
\bottomrule
\end{tabular}
\caption{\textbf{Overall TPR and FDR.} This table displays the average (standard deviation) overall true positive rate (TPR) and false discovery rate (FDR) when using independent Bernoulli priors for variable selection or the Markov random field (MRF) prior.}
\label{tab:performance}
\end{table}

\section{Analysis of MLVS and MBS}
\subsection{Data}

HPFS is an ongoing prospective cohort study that began in 1986 and includes 51,529 US male health professionals \citep{Li_Wang_Li_Ivey_Wilkinson_Wang_Li_Liu_Eliassen_Chan_etal._2022, Li_Wang_Satija_Ivey_Li_Wilkinson_Li_Baden_Chan_Huttenhower_etal._2021}. NHSII is an ongoing prospective cohort study that began in 1989 and includes 116,429 female registered nurses \citep{Huang_Trudel-Fitzgerald_Poole_Sawyer_Kubzansky_Hankinson_Okereke_Tworoger_2019, Ke_Guimond_Tworoger_Huang_Chan_Liu_Kubzansky_2023}. In both studies, biennial measurements are collected on lifestyle factors such as diet, exercise, sleep, physical and mental health, and medication use \citep{Rimm_Giovannucci_Willett_Colditz_Ascherio_Rosner_Stampfer_1991, Bao_Bertoia_Lenart_Stampfer_Willett_Speizer_Chavarro_2016}. The present analysis centers on the substudies MLVS and MBS. Briefly, MLVS ran from 2011 to 2013 and included 700 randomly selected participants from HPFS aged 52–81. MBS was a 2013-2014 substudy of NHSII that included 233 randomly selected women aged 49-67 \citep{Huang_Trudel-Fitzgerald_Poole_Sawyer_Kubzansky_Hankinson_Okereke_Tworoger_2019}. Participants of either substudy were free of coronary heart disease, stroke, cancer, and major neurological disease at the time of enrollment. Both substudies involved the collection of two blood samples and two pairs of stool samples at six-month intervals for the purpose of metabolome and microbiome profiling. For consistency with recent analyses on the relationship between metabolites and microbiome composition \citep{Li_Li_Ivey_Wang_Wilkinson_Franke_Lee_Chan_Huttenhower_Hu_etal._2022,  Li_Wang_Li_Ivey_Wilkinson_Wang_Li_Liu_Eliassen_Chan_etal._2022}, we restricted analysis to the 203 participants in MBS and 287 participants in MLVS for whom microbiome and metabolite data are both available. Further details on metabolite measurement are given by \citet{Manghi_Bhosle_Wang_Marconi_Selma-Royo_Ricci_Asnicar_Golzato_Ma_Hang_etal._2024}.

We quantified diet based on a validated food frequency questionnaire \citep{Willett_Sampson_Stampfer_Rosner_Bain_Witschi_Hennekens_Speizer_1985, Gu_Wang_Sampson_Barnett_Rimm_Stampfer_Djousse_Rosner_Willett_2024, Yuan_Spiegelman_Rimm_Rosner_Stampfer_Barnett_Chavarro_Rood_Harnack_Sampson_etal._2018} that measured how often respondents had eaten specific foods over the previous 12 months, from ``never or less than once/month'' to ``$\geq 6$ times/day.'' The questionnaire contained $>130$ items and was administered every 4 years beginning in 1986 for HPFS and in 1991 for NHSII. To summarize each person's long-term consumption habits for each survey item, we averaged over the survey responses that preceded blood sample collection. Finally, we aggregated the responses into 30 variables representing the average daily servings of specific foods types, listed in \cref{sec:data_analysis}.

\subsection{Analysis}
\label{sec:data_analysis}

We analyzed the 244 metabolites with $< 80\%$ PMVs in both cohorts. For each metabolite in each cohort, we fit four regression models that assumed normal or skew-normal errors and used independent Bernoulli priors or the MRF prior. All models adjusted for age, BMI, and race (White/non-White) as confounding variables; continuous variables were $z$-standardized. To place metabolites on a comparable scale, we replaced PMVs with half-minimum values, standardized, and re-introduced the PMVs into the data before analysis.

\subsubsection{Hyperparameters and implementation}

We chose $\psi$, $\rho_0$, and $\rho_1$ on a metabolite-specific basis by the approach described in \cref{sec:sim_analysis}. We set $\nu^2$ to the empirical variance of the single-predictor regression coefficients, a common practice in Bayesian variable selection \citep{Li_Zhang_2010}. These models were adjusted for the same confounding variables as in the main analyses and used the normalized, imputed metabolites as responses. We used standard weakly informative priors for the intercept, skewness, and confounder effects, setting $\nu^2_0=\nu^2_d=\lambda_1^2=\lambda_2^2=\lambda_3^2=10^2$  \citep{Wadsworth_Argiento_Guindani_Galloway-Pena_Shelburne_Vannucci_2017}. We set $(\xi_0,\sigma_0^2)=(3,1)$ so that $\sigma^{-2}$, the precision, would have prior mean 1 and variance 1.5. We let $\omega=\text{logit}(0.05)$ to promote sparsity, and we searched for $\eta$ over the range $0.01, 0.02,\dots,1.00$.

For the MCMC, we drew 300,000 posterior samples following a burn-in period of 30,000 samples. We thinned the samples by 95\% to reduce autocorrelation. Convergence was evaluated by running three MCMC chains and visually comparing the trace plots and posterior means of the parameters. To account for testing associations in multiple models, we declared signals to be significant if their PIP exceeded a 5\% Bayesian FDR threshold over all combinations of metabolites and food intake variables. 

\subsubsection{Informative variable selection prior}
\label{sec:diet_R}
\begin{figure}[ht]
    \centering
    \includegraphics[width=6in]{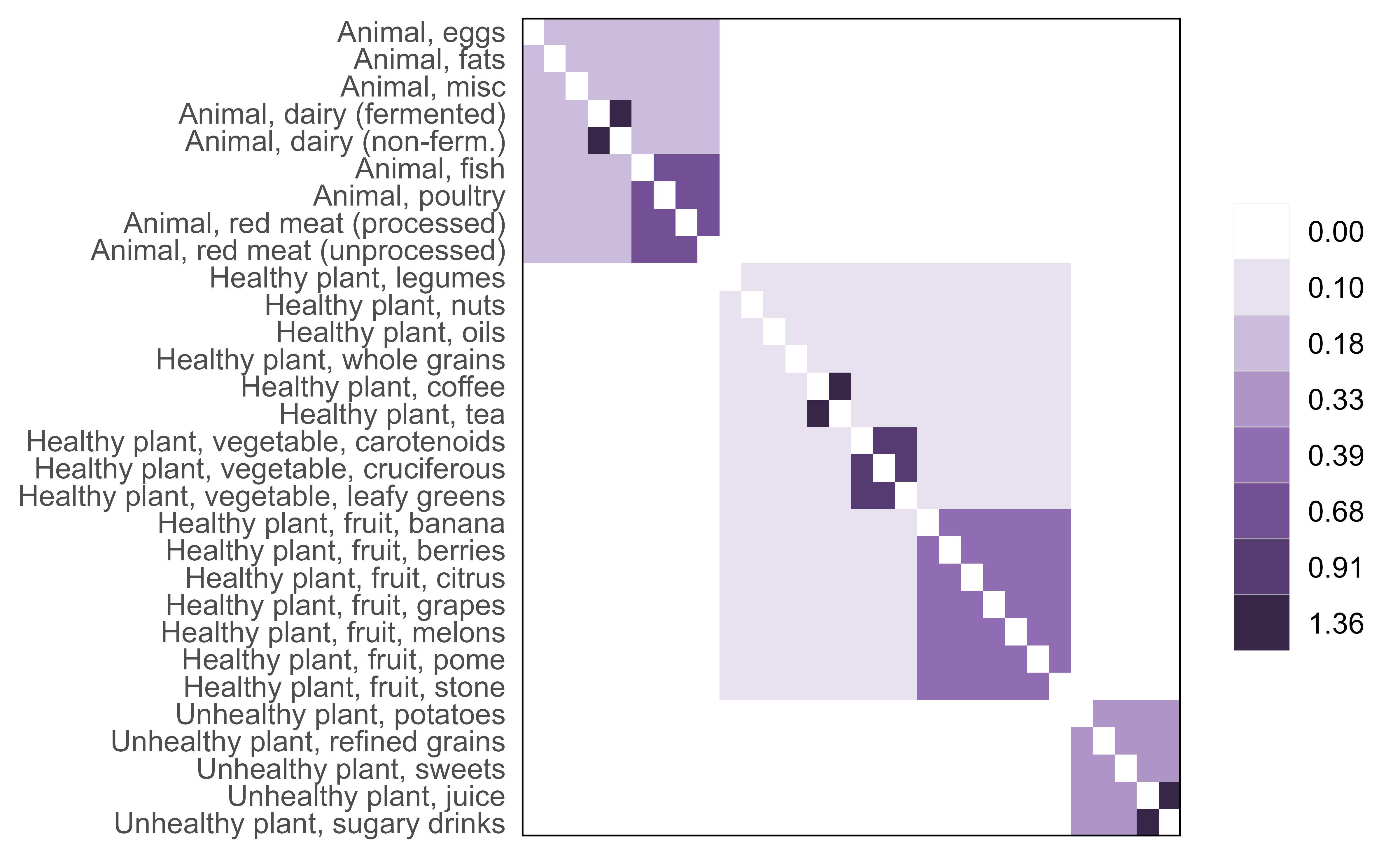} 
    \caption{\textbf{Dietary relationship matrix ($\bm{R}$}). The figure depicts empirical pairwise relationships among the 30 dietary intake variables considered for variable selection. Variables were grouped hierarchically based on their nutritional properties. $\bm{R}$ was incorporated into the variable selection scheme via the Markov random field prior.}
    \label{fig:dietary_R}    
\end{figure}

To construct $\bm{R}$, we used previously defined groupings based on the nutritional and biological relationships among food items.  \citet{Satija_Bhupathiraju_Rimm_Spiegelman_Chiuve_Borgi_Willett_Manson_Sun_Hu_2016} classified the food items in HPFS and NHSII as healthy plant-based, unhealthy plant-based, or animal-based foods, given their established associations with risk of type II diabetes and related health outcomes. Within these broad classes, we defined subcategories: fruits, vegetables, and healthy beverages (within healthy plant-based), unhealthy beverages (within unhealthy plant-based), and dairy and meat/fish/poultry (within animal-based). We derived $\bm{R}$ was derived from this hierarchical structure using the approached described in \cref{sec:data_gen} (\cref{fig:dietary_R}),

\subsubsection{Model Validation}
We evaluated the quality of the model fit in two complementary ways. First, for three metabolites with the strongest signals, we visually inspected the posterior predictive distribution of $Y_i$ by overlaying it on the empirical distribution. Second, for all metabolites, we compared the normal and skew-normal error models by estimating their expected log posterior predictive density (ELPD), quantifying the out-of-sample predictive ability \citep{Gelman_Vehtari_Simpson_Margossian_Carpenter_Yao_Kennedy_Gabry_Bürkner_Modrák_2020}. Models with higher ELPD values capture the true distribution of the outcome more accurately.  The ELPD can be estimated within-sample by using importance sampling (IS) or the Watanabe-Akaike Information Criterion (WAIC), based on the log-likelihood of posterior draws \citep{Gelfand_Dey_Chang_1992,Watanabe_2010}. For each combination of cohort, error distribution, and prior specification,  we summed log-likelihoods across individual metabolite models to compute a single multivariate log-likelihood. We used this to estimate the ELPD using both IS and WAIC, summarizing the quality of fit over all metabolites.

\subsection{Results}

\begin{figure}[htbp]
    \centering
    \includegraphics[width=6.5in]{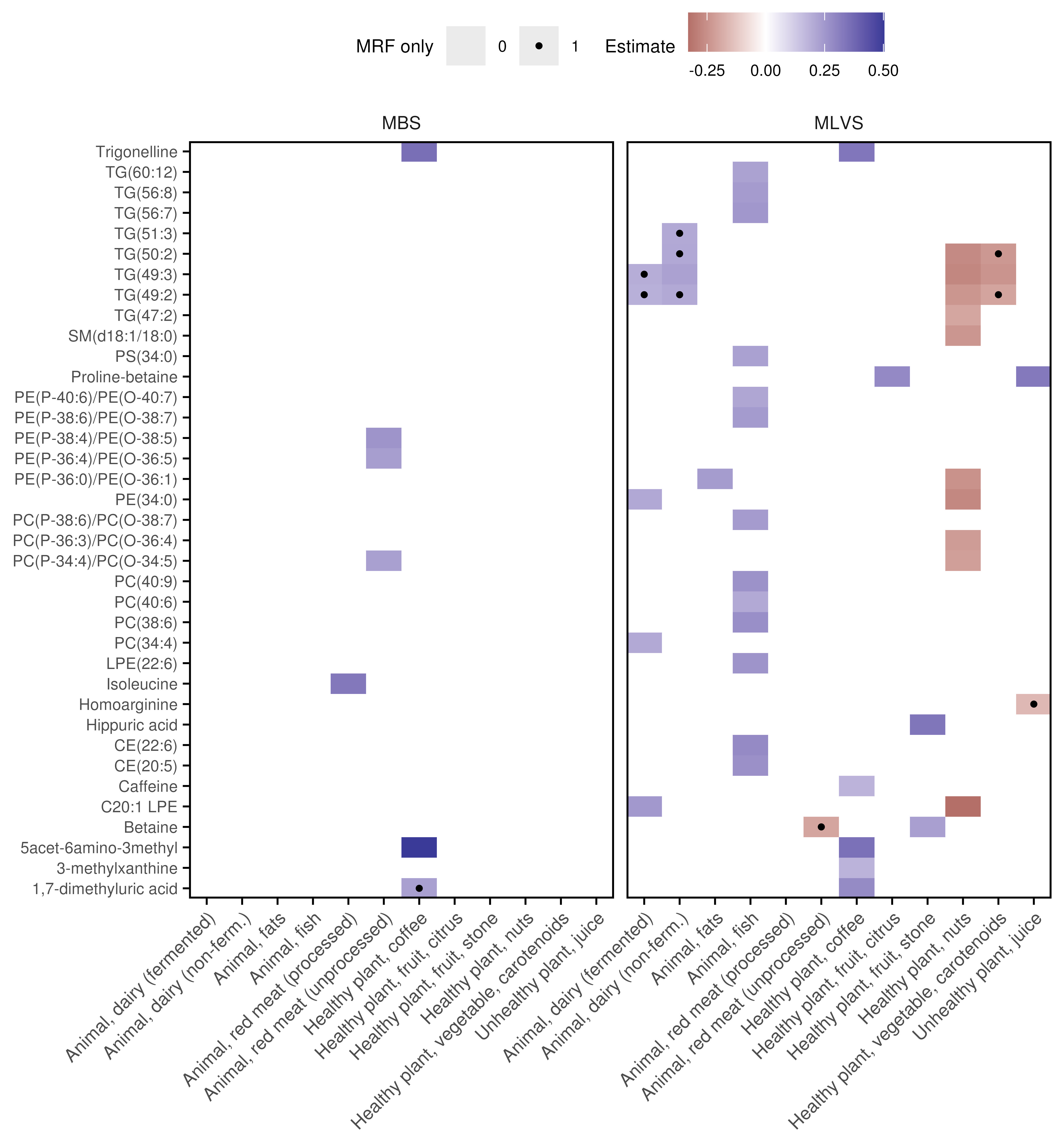} 
    \caption{\textbf{Metabolite-dietary associations identified by SNCM model}. The figure depicts the associations detected in MBS or MLVS when fitting the skew-normal censored mixture (SNCM) model with a Markov random field prior (MRF). Associations with a black dot were not identified when using independent Bernoulli priors, only the MRF prior. Coefficient estimates are on the scale of 1 standard deviation of the response and predictor.}
    \label{fig:variable_chosen}
\end{figure}

Using the SNCM model with an MRF prior, we identified 37 unique metabolites associated with at least one food item in either the MBS or MLVS (\cref{fig:variable_chosen}). In total, there were 54 food–metabolite associations across the two cohorts. Twelve food items were associated with at least one metabolite; fish (13 associations), nuts (10), coffee (8), and fermented dairy products (5) produced the most signals. Only seven of the 54 total associations were identified in the MBS. Fifteen of the 54 associations were negative, including every signal for the dietary factors nuts, red meat, and carotenoid vegetables. 

In the MLVS, analyses using the MRF prior identified nine associations that were not detected when using independent Bernoulli priors. Five of these discrepancies were positive associations between fermented and/or non-fermented dairy products and the metabolites C51:3 triacylglycerols (TG), C50:2 TG, C39:3 TG, and C49:2 TG. The other four were negative associations between carotenoid vegetables and C50:2 TG and C49:2 TG, between unhealthy fruit juice and hippuric acid, and between unprocessed red meat and betaine. In the MBS, analyses using the MRF prior, but not the independent priors, identified a positive link between coffee consumption and 1,7-dimethyluric acid. There were no associations in either study that were detected only using independent priors.

\subsubsection{Validation}
Across all metabolites and both cohorts, models with skew-normal errors offered better fit than their normal-error counterparts (\cref{tab:validation}). Using the MRF prior instead of the independent prior resulted in improved model fit, though the gain was less pronounced than the improvement achieved by switching from a normal to a skew-normal error model. 

\begin{figure}[ht!]
    \centering
    \includegraphics[width=6.5in]{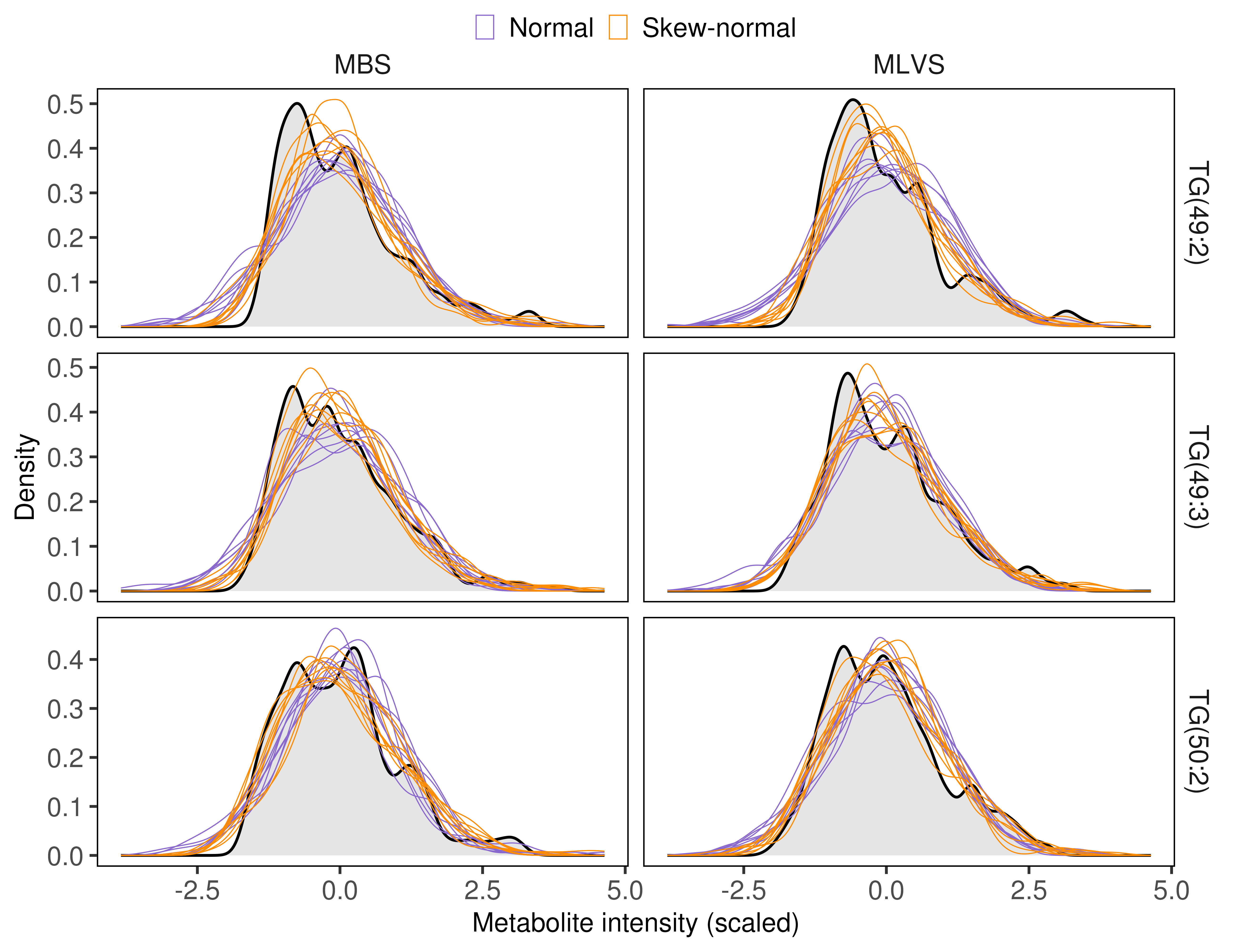} 
    \caption{\textbf{Posterior predictive distributions from the normal and skew-normal models}. The shaded region depicts the empirical density of a metabolite in either the Mind Body Study (MBS) or Men's Lifestyle Validation Study (MLVS). For models assuming either a normal (purple) or skew-normal (orange) error term, we show 8 densities sampled from the joint posterior distribution of the model parameters, approximating the posterior predictive distribution.}
    \label{fig:post_pred}
\end{figure}

TG(49:2), TG(49:3), and TG(50:2) were the three metabolites associated with the largest number of food items. Based on the posterior predictive distributions of these metabolites among MRF-based models (\cref{fig:post_pred}), the skew-normal model (orange curves) more closely matched the empirical density than the normal model (purple curves) in most cases. While the orange curves tended to be centered properly around the peak of the empirical density, the purple curves overestimated the true mode of the density due to skewness, demonstrating the skew-normal model's superior fit. These patterns were subtle in some cases but were pronounced for TG(49:2).

\begin{table}[ht]
\centering
\begin{tabular}{llrrrr}
\toprule
\textbf{Cohort} & \textbf{Selection Prior} & \multicolumn{2}{c}{$\hat{\text{ELPD}}_\text{IS}$} & \multicolumn{2}{c}{$\hat{\text{ELPD}}_\text{WAIC}$} \\
\cmidrule(lr){3-4} \cmidrule(lr){5-6}
& & \textbf{Normal} & \textbf{Skew-normal} & \textbf{Normal} & \textbf{Skew-normal} \\
\midrule
MBS & Independent & -69858.84 & -63919.18 & -70246.73 & -64084.26 \\ 
MBS & MRF & -69813.74 & -63869.16 & -70241.76 & -64054.69 \\ 
MLVS & Independent & -98789.91 & -88961.87 & -99383.80 & -89196.88 \\ 
MLVS & MRF & -98673.37 & -88849.07 & -99287.46 & -89102.81 \\ 
\bottomrule
\end{tabular}
\caption{\textbf{Comparison of model fit by estimated ELPD}. Each column displays the estimated expected log posterior density (ELPD) for the normal or skew-normal censored mixtured model. Datasets were derived from the Mind Body Study (MBS) or Men's Lifestyle Validation Study (MLVS). Models used  independent Bernoulli priors or the Markov random field (prior) for variable selection. We estimated ELPD using importance sampling (IS) and the Watanabe-Akaike information criterion (wAIC). Higher values indicate better model fit.}
\label{tab:validation}
\end{table}

\section{Discussion}
To characterize the metabolomic signatures of 30 food intake variables in two cohorts of healthcare professionals, we developed a novel Bayesian variable selection framework that combines a SNCM model with a substantively informed MRF prior. A broad spectrum of simulation experiments demonstrated that the SNCM model effectively identifies true predictors of metabolite abundance in the presence of high PMV rates, heavy skewness, and inter-predictor correlations. Using the MRF prior to incorporate external information led to substantial improvements in overall TPRs without inflating FDRs, while greatly improving the individual TPR of predictors with weak signals. In simulations where the external knowledge was uninformative of the predictive variables, the MRF prior had little effect on TPR or FDR, suggesting there is no intrinsic penalty if $\bm{R}$ is misspecified. On real data, the MRF-based SNCM model identified more dietary predictors than alternatives while offering superior model fit. The framework can implemented using the R package \texttt{multimetab}, available on \textcolor{blue}{\href{https://github.com/dclarkboucher/multimetab}{GitHub}}. While designed for metabolomics, the proposed method should prove useful for other 'omics response types, such as proteomics or viral load data, which share many of the same analytical challenges \citep{Schwämmle_Hagensen_Rogowska-Wrzesinska_Jensen_2020,Dagne_Huang_2013}. 

 Several findings align with known metabolic pathways. 5-acetylamino-6-amino-3-methyluracil and 1,7-dimethyluric acid, two established caffeine-derived metabolites \citep{Batista-da-Silva_Limirio_DeOliveira_2024}, were positively associated with coffee intake in both cohorts. PC(40:6) and PE(P-40:6)/PE(0-40:7), which were associated with fish consumption in the MLVS, contain substituents such as docosahexaenoic acid (DHA) which is derived from fish oils \citep{Vauzour_Scholey_White_Cohen_Cassidy_Gillings_Irvine_Kay_Kim_King_etal._2023}. In the MLVS, C49:2 TG and C49:3 TG were selected for fermented and/or non-fermented dairy intake by the MRF prior; both  triacylglycerols contain fatty acyl (pentadecanoic acyl chain, 15:0) chains that serve as potential biomarkers of dairy intake \citep{Abdullah_Cyr_Lépine_Labonté_Couture_Jones_Lamarche_2015}. Meanwhile, consumption of unhealthy plant juice was negatively associated with the amino acid homoarginine, and low homoarginine is correlated with worse cardiovascular health \citep{Tsikas_2023}. Several of these signals were not detected when using independent Bernoulli priors, and it is plausible to speculate that the nutritionally informed model identified some true metabolomic signatures that were missed by the model with indpendent priors. 

A key strength of the proposed framework is its sensitivity and practical utility in settings with high-dimensional and correlated predictors, as evidenced by the simulation experiments with $p=300$. Such scenarios arise often in omic and multi-omic applications and pose substantial challenges for standard statistical methods. This work demonstrates the value of adopting sophisticated statistical frameworks---such as the MRF prior---that enhance signal detection  by exploiting domain knowledge about the relationships under study. 

The framework's strong performance lays the foundation for statistical and applied extensions to more complex types of covariates, such as microbial taxa. Microbiome studies often involve hundreds of taxa simultaneously, with measurements that are noisy, zero-inflated, and compositional \citep{Gloor_Macklaim_Pawlowsky-Glahn_Egozcue_2017}. Although there are existing methods for microbiome covariates incorporating MRF priors to exploit taxonomic or phylogenetic structure, such approaches are not engineered for 'omics responses generated by mass-spectrometry workflows \citep{Zhang_Shi_Jenq_Do_Peterson_2021}. We therefore plan to extend the proposed framework to incorporate microbial covariates while explicitly accounting for these added features. 

Explicitly modeling the skewness of the metabolites via the skew-normal distribution circumvented the need for ad hoc logarithmic (or other) transformations and yielded uniformly better fit than a normal model. One could generalize this framework by instead adopting a skewed t model, potentially capturing skewed and heavy-tailed metabolite distributions even more effectively. The main drawback of this approach would be the additional sampling of the degrees of freedom parameter \citep{Sahu_Dey_Branco_2003}.

A potential methodological limitation of the proposed framework is that it treats $\rho$, the biological PMV rate, as the same for all individuals. Some analysis methods for metabolite responses make the same simplifying assumption \citep{Shah_Brock_Gaskins_2019}, whereas others allow $\rho$ to vary across subjects \citep{Huang_Lane_Fan_Higashi_Weiss_Yin_Wang_2020}. An ambitious extension would be to model $\rho$ as a function of $\bm{C}_i$ and $\bm{X}_i$, inferring relationships between the biological PMV rate and the primary predictors. However, this approach could present substantial computational challenges and identifiability issues due to the complexity of performing variable selection in zero-inflated models---especially if the biological PMV rate is low \citep{Lee_Coull_Moscicki_Paster_Starr_2020}. 

When implementing variable selection, we adopted the spike-and-slab prior formulation by \citet{Kuo_Mallick_1998} instead of the more commonly used variance-mixture form, $\beta_j\sim N(0,\gamma_j\nu^2)$. Early versions of the proposed method implemented the latter formulation using stochastic search variable selection (SSVS), which requires a Metropolis-Hastings algorithm since Gibbs sampling is not feasible \citep{George_McCulloch_1997}. However, the SVSS struggled in multiple facets of MCMC convergence: it required many more posterior samples to reach the target distribution, exhibited high autocorrelation between successive draws, and produced unstable posterior means of $\bm{\gamma}$ across chains. While a rigorous comparison of these implementations is beyond the current scope, these observations suggest that for high-dimensional models with complex responses, researchers should carefully assess the stability of SVSS-style sampling schemes and consider alternatives  when appropriate. 

Identifying dietary predictors of metabolite abundance is essential for understanding the nutritional foundations of human health. Our work sheds light on such mechanisms by presenting a general, flexible framework for performing variable selection with metabolite responses. The framework was greatly enhanced by incorporating external nutritional information via the MRF prior, offering stronger performance in simulations and detecting more food-metabolite associations on the MLVS data. The framework's efficacy in high-dimensional covariate settings paves the way for multi-'omics analyses that can contribute to an integrative understanding of biological pathways. As the breadth and complexity of 'omics-based applications continues to expand, statistical frameworks that can robustly handle complex types of data will be essential for making transformative biological insights.

\section*{Acknowledgements}
This work was supported by the National Institute of General Medical Sciences (R01GM126257). The Health Professionals Follow-Up study and Nurse’s Health Study were supported by the National Cancer Institute (U01CA167552, 28U01CA176726).
\section*{Disclosure statement}\label{disclosure-statement}

The authors have declared that there are no conflicts of interest. 

\section*{Data Availability Statement}\label{data-availability-statement}

The data used in this study cannot be shared publicly for confidentiality reasons. To request data access, follow the instructions provided on the \textcolor{blue}{\href{https://www.nurseshealthstudy.org/researchers}{Nurse's Health Study}} or \textcolor{blue}{\href{https://www.hsph.harvard.edu/hpfs/for-collaborators/}{Health Professionals Follow-up Study}} websites.

\phantomsection\label{supplementary-material}
\bigskip

\begin{center}

{\large\bf SUPPLEMENTARY MATERIAL}

\end{center}

\begin{description}
\item[Supplementary materials] A supplementary file containing additional figures, results and details about the MCMC scheme. (PDF)
\end{description}

\bibliography{bibliography.bib}

\end{document}